\documentclass[twoside]{article}
\usepackage{fleqn,espcrc2}
\input epsf
\begin{document}

\author{V. Ta Phuoc, A. Ruyter, R. de Sousa, L. Ammor, E. Olive, J.C. Soret
\address{Laboratoire d'Electrodynamique des Mat\'{e}riaux Avanc\'{e}s.\\ Universit\'{e} 
F.Rabelais - UFR Sciences - Parc de Grandmont.\\37200 Tours- France.}}
\title{Effect of field tilting on the vortices in irradiated Bi-$2212$.}
\begin{abstract}
We report on transport measurements in a Bi-$2212$ single crystal with
columnar defects parallel to the c-axis. The tilt of the magnetic field away
from the direction of the tracks is studied for filling factors $f=B_z/B_\phi
<1$. Near the Bose Glass transition temperature $T_{BG}$, the angular
scaling laws are verified and we find the field independent 
critical exponents $\nu ^{\prime}=1.1$ and $z^{\prime }=5.30$. 
Finally, above $H_{\perp C}$ we evidence the
signature of a smetic-A like vortex phase. These experimental results
provide support for the Bose Glass theory.  
\end{abstract}
\maketitle
Columnar defects have been introduced in HTCS in order to avoid dissipation
due to vortex motion. In the case of parallel tracks Nelson and Vinokur\cite
{Nelson1993} have predicted a transition between a so-called Bose Glass (BG) 
and vortex liquid, at a critical temperature $T=T_{BG}$.
An important result is that field tilting does not
destabilize the BG phase. Above a threshold transverse field $H_{\bot C}(T)$
perpendicular to the columnar defects (CD), the flux lines accomodate
simultaneously to CD and to the transverse field direction, defining a
smectic-A like phase as proposed by Hwa and Nelson\cite{Hwa1993} .
Increasing further $H_{\bot }$ leads to a vortex-liquid state 
\cite{Hatano1996}. Recently, Grigera \textit{et al}. \cite{Grigera1998} have
evidenced a threshold transverse field $H_{\bot C}(T)$ with resistivity
measurements performed on a twinned Y-123 single crystal.

In this paper, we present results obtained on a Bi-2212 single crystal, irradiated 
with CD parallel to the c-axis with 5.8 GeV $Pb$
ions in GANIL (Caen- France). The CD density corresponds to a matching
field $B_\phi =0.75$ T.
Isothermal I-V curves have been obtained varying the tilt angle between
the magnetic field and the c-axis of the crystal. Measurements were performed
keeping the $B_z$ component constant, where the z-axis is along the c-axis
of the sample.
In the insert of
Fig.\ 1, the log-log plot of isothermal ohmic resistance versus $H_{\bot
}/H_z$ is displayed above and below $T_{BG}$. 
For $T>T_{BG}$, ohmic behaviour is detected, within the experimental sensitivity, even 
for vanishing transverse field. 
In contrast, for $T<T_{BG}$ the ohmic resistance goes to zero at some critical tilt showing the
existence of a critical transverse magnetic field $H_{\bot C}(T)$. 
In the BG theory, scaling functions $f_{+}$ and $f_{-}$ are predicted
to describe the resistivity in the presence of a tilted magnetic field, above and below $T_{BG}$,
respectively. The resistance then reads: 
\begin{equation}
R\sim |t|^{\nu ^{\prime }(z^{\prime }-2)}f_{\pm }\left( (H_{\perp
}/H_z)|t|^{-3\nu ^{\prime }}\right).
\end{equation}
\begin{figure}
\epsfxsize=7. cm 
\epsfbox{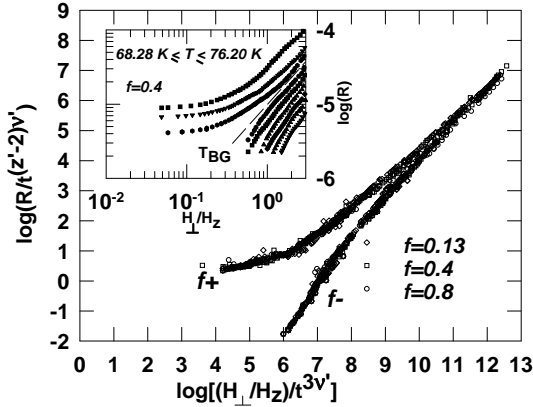}
\caption{Scaling properties according to Eq. 1 for different filling fractions $f=0.13$, $f=0.4$, and $f=0.8$.
The field independent critical exponents we obtain are $\nu^{\prime}=1.1
\pm 0.1$ and $z^{\prime}=5.30\pm 0.05$. The insert displays a log-log plot
of isothermal ohmic resistance versus $H_{\perp}/H_{z}$ above and below $T_{BG}$ for $f=0.4$.}
\end{figure}
where $t=(T-T_{BG})/T_{BG}$ is the reduced temperature, and $\nu ^{\prime }$
and $z^{\prime }$ are critical exponents. Fig. 1 displays such scaling
properties. We find that the scaling functions $f_{+}$ and $f_{-}$ 
collapse for all the filling fractions $f=B_z/B_\phi <1$ investigated in our experiment. 
Moreover, we obtain field independent critical exponents, $\nu ^{\prime }=1.1\pm
0.1$ and $z^{\prime }=5.30\pm 0.05$. These results
are in good agreement with the BG theory \cite{Nelson1993} and other
experimental results \cite{Grigera1998,Soret1998,TaPhuoc1997}.

We shall now consider the case $T<T_{BG}$. For $H_{\bot }>H_{\bot C}(T)$,
the vortex motion is mediated by kinks aligned in chains in the direction of
the transverse magnetic field $H_{\bot }$ \cite{Hwa1993}, in such a way that the chain density
is directly related to the linear resistance: 
\begin{equation}
R\propto n_{chain}\sim (H_{\perp }-H_{\perp C}(T))^{3/2}.
\end{equation}
\begin{figure}
\epsfxsize=7.2 cm 
\epsfbox{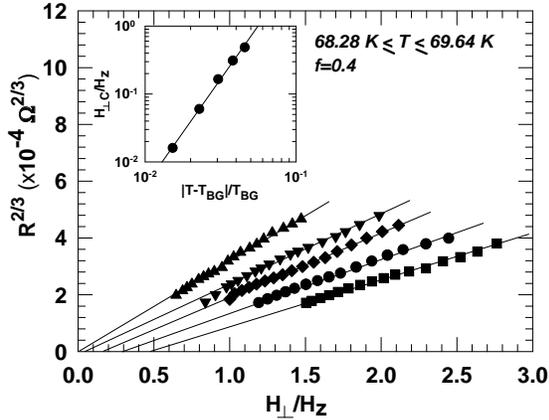}
\caption{Plot of $R^{2/3}$ versus $H_{\perp}/H_{z}$. The data exhibit linear 
behaviour as expected in Eq. 2. The intersections of the lines with the abscissa
axis give the critical transverse fields for different temperatures.
The insert displays the transverse critical field 
$H_{\perp c}/H_{z}$ versus the reduced temperature. The slope gives directly the
critical exponent value $\nu^{\prime}=1.1\pm 0.1$. }
\end{figure}
Scaling arguments
lead to a critical transverse magnetic field $H_{\perp C}(T)$ vanishing as 
$T\rightarrow T_{BG}^{-}$ as: 
\begin{equation}
H_{\perp C}(T)\sim |t|^{3\nu ^{\prime }}.
\end{equation}

Figure 2 displays a plot of $R^{\frac 23}$ versus $H_{\perp }/H_z$. 
The solid lines are successfull fits of Eq. 2 to data. This result does support
the existence of a smectic-A behaviour. The intersection of a linear fit
with the abscissa-axis directly gives the critical transverse magnetic field $%
H_{\perp C}(T)$ at a given temperature. The insert of Fig. 2 shows a log-log
plot of $H_{\perp C}(T)$ thus obtained versus the reduced temperature 
$t=(T-T_{BG})/T_{BG}$. The solid line represents a least-square fit of Eq. 3
to data. According to Eq. 3, we find therefrom the critical exponent value $\nu ^{\prime
}=1.1\pm 0.1$. Note that this value is consistent with the one we found above by another way, as
predicted by Nelson and Vinokur \cite{Nelson1993}.

In conclusion, we have investigated the Bose Glass phase transition on
an irradiated Bi-2212 single crystal versus both the filling fraction $%
f=B_z/B_\phi <1$ and the magnetic field tilt. We have shown that the Bose
Glass transition in the presence of a tilted field verifies the scaling
rules predicted by Nelson and Vinokur with both field independent 
scaling functions and critical
exponents $\nu ^{\prime }=1.1$ and $z^{\prime }=5.30$. A smectic-A like
behaviour has been evidenced. Finally, the critical tilt $H_{\perp C}/H_z$ ,
separating this phase from the Bose Glass one, has been found to vary as $%
H_{\perp C}/H_z\sim |t|^{3\nu ^{\prime }}$ in agreement with the theorical
expectation of a sharp cusp in the $T-H_{\perp }/H_z$ phase diagram.

\end{document}